# COVID-19 Vaccine Acceptance in the US and UK in the Early Phase of the Pandemic: AI-Generated Vaccines Hesitancy for Minors, and the Role of Governments[*]

*Gabriel Lima*[1], *Meeyoung Cha*[2], *Chiyoung Cha*[3], *Hyeyoung Hwang*[4]

## Abstract

This study presents survey results of the public's willingness to get vaccinated against COVID-19 during an early phase of the pandemic and examines factors that could influence vaccine acceptance based on a between-subjects design. A representative quota sample of 572 adults in the US and UK participated in an online survey. First, the participants' medical use tendencies and initial vaccine acceptance were assessed; then, short vignettes were provided to evaluate their changes in attitude towards COVID-19 vaccines. For data analysis, ANOVA and post hoc pairwise comparisons were used. The participants were more reluctant to vaccinate their children than themselves and the elderly. The use of artificial intelligence (AI) in vaccine development did not influence vaccine acceptance. Vignettes that explicitly stated the high effectiveness of vaccines led to an increase in vaccine acceptance. Our study suggests public policies emphasizing the vaccine effectiveness against the virus could lead to higher vaccination rates. We also discuss the public's expectations of governments concerning vaccine safety and present a series of implications based on our findings.

## 1. Introduction

The coronavirus disease 2019 (COVID-19) pandemic has been a traumatic event that has caused intense fear, helplessness, and suffering (Cho, Park, 2020). Fortunately, the global race to successfully deliver highly effective vaccines could be the ultimate solution to the global pandemic. Recent phase III trials have achieved efficacy levels that could change the COVID-19 landscape (Polack et al., 2020; Voysey et al., 2020). Countries with high vaccination rates, such as Israel, exemplify how widespread vaccination can prevent mass hospitalizations and deaths (Rossman et al., 2021). However, attempts to accelerate vaccination have been delayed by reports of side effects that have led countries around the world to stop immunization programs for further investigation (Mueller, 2021).

[*]This work was supported by the Institute for Basic Science (IBS-R029-C2) & the National Research Foundation of Korea (NRF) grant funded by the Korea government (MSIT) (No. 2021R1A2C2008166).
[1]Graduate Student, School of Computing, KAIST, 291 Daehak-ro, Yuseong-gu, Daejeon, 34141, Republic of Korea. E-mail : gabriel.lima@kaist.ac.kr
[2]Chief Investigator & Associate professor, Institute for Basic Science (IBS) & School of Computing, KAIST, 55 Expo-ro, Yuseong-gu, Daejeon, 34126, Republic of Korea. E-mail : mcha@ibs.re.kr
[3](Corresponding author) Associate professor, College of Nursing & Ewha Research Institute of Nursing Science, Ewha Womans University, 52, Ewhayeodae-gil, Seodaemun-gu, Seoul, 03760 Republic of Korea. E-mail : chiyoung@ewha.ac.kr
[4]Registered Nurse, Samsung Medical Center, 81 Irwon-Ro Gangnam-gu. Seoul, 06351, Korea. E-mail : hwanghy.grace@gmail.com
[Received 6 April 2021; Revised 7 May 2021, 28 May 2021; Accepted 1 June 2021]

The successful development of COVID-19 vaccines has raised a series of novel challenges for controlling the novel coronavirus. Researchers, policymakers, and healthcare workers are rushing to ensure that vaccines are distributed efficiently and equitably. Challenges to delivering coronavirus vaccines range from the scale of manufacturing to the coordination of logistics worldwide (Kim et al., 2021). Even if these difficulties are resolved and vaccines become available to the global population, vaccine hesitancy may still hinder the eradication of the novel coronavirus. Studies indicate that vaccine hesitancy could be an obstacle to herd immunity (Corse, 2021; Cohen, 2021). To what extent global citizens are willing to accept COVID-19 vaccines is an important question whose answer will influence future health care decisions. Refusal to be vaccinated against COVID-19 could lead to a longer-standing pandemic and deaths that could have been prevented by widespread vaccination.

New technologies are being used in the development of future vaccines. For example, artificial intelligence (AI) has been adopted to shorten the vaccine development process by reviewing possible matches for already available drugs in the market, repurposing them for other diseases, and developing new ones (Yassine, Shah, 2020). However, the use of AI has raised various moral and legal questions regarding responsibility (Coeckelbergh, 2020; Gunkel, 2017). Deploying these systems in life-or-death scenarios such as medical situations could raise the stakes and lead to more complicated issues.

The current research investigates the hindering factors in the public acceptance of COVID-19 vaccines, including the use of AI systems in their development. Understanding the public's willingness regarding this subject is critical as vaccines are often the ultimate solution to reducing the disease burden. Such an understanding could also allow researchers and policymakers to better prepare for vaccine hesitancy due to new technologies. Examining AI as one of these hindering factors is important because introducing AI in biotechnology for vaccine development requires re-evaluating people's perception of newly developed vaccines.

The value of our study is that it captures public sentiment during the early stage of the pandemic in mid-April 2020, the period during which initial perceptions of vaccines were set. Our study reveals people report different levels of hesitancy depending on the recipient and that news reports regarding a vaccine's effectiveness and side effects can significantly influence vaccine uptake. Crucially, whether AI is used in the vaccine development process did not seem to influence vaccine acceptance. Finally, we address several ethical issues related to AI-developed vaccines and observe that governments are expected to take a leading role in ensuring vaccine safety. Hence, the findings presented in this research hold a historic value that could improve the way public perceptions and concerns regarding a novel disease are addressed.

## 2. Background

### 2.1. COVID-19 and Vaccination

COVID-19 is a newly discovered infectious disease with respiratory symptoms caused by SARS-CoV-2 (Wang et al., 2020). While estimates of the disease's reach and death rate constantly change during its continuous spread, early data suggest that certain groups, such as elderly individuals and patients with pre-existing medical conditions, are more likely to become severely ill from COVID-19. For instance, researchers have estimated a 20 times higher death rate among those aged over 60 years than their younger counterparts (Verity et al., 2020).

The World Health Organization (WHO) and national health organizations worldwide have been promoting public policies, such as quarantines, that have led some countries to lock down their entire population. Protests around the world emerged to fight against these governmental decisions (Andone, 2020). These public policies, which often conflict with human rights, are necessary to contain the disease and are justified by the disease's threat to public health (Park, Kim, Park, 2020). Countries that faced the novel coronavirus earlier in the pandemic started to loosen restrictions, such as enforced lockdowns, while promoting more lenient policies, e.g., social distancing among their citizens. Doing so, however, has led to second waves of infections in some countries (Xu, Li, 2020).

The COVID-19 pandemic has also created a so-called infodemic (Earnshaw, Katz, 2020), in which a plethora of false information regarding the virus is rapidly being shared online. Worryingly, low-income countries seem to be the most susceptible to online misinformation (Cha et al., 2021). Studies indicate that misinformation about Ebola has negatively influenced vaccine acceptance during previous Ebola epidemics (Vinck et al., 2019). The WHO has also reported that misinformation has led to unfounded controversies about the safety of the HPV vaccine (WHO, 2019). As knowledge of HPV and its vaccine becomes significant factors influencing vaccine decisions (Kang, Jun, 2013; Choi, Park, 2012; Lee, Kang, Lee, 2010), providing accurate information could increase vaccine acceptance. These findings suggest that the COVID-19 infodemic might contribute to vaccination hesitancy and refusal, an alarming note in the current pandemic.

### 2.2. Vaccination and Artificial Intelligence

Historically, vaccination has been successful in controlling diseases (Orenstein, Ahmed,

2017). The WHO estimated that more than 17.1 million lives had been saved from measles due to widespread vaccination from 2000 to 2015 (WHO, 2015). Philanthropists have invested millions of dollars in developing new vaccines to combat diseases that still affect certain areas of the world, such as malaria in Africa. However, the WHO ranked vaccine hesitancy among the top 10 health threats in 2019. Anti-vaccine movements still contribute to an increase in vaccine-preventable outbreaks and epidemics by promoting hesitancy. These harmful fronts sway vaccine acceptance and are often influenced by conspiracy theories (Jolley, Douglas, 2014), lack of trust in the health system, past experiences with vaccination, and other factors (Dubé, Vivion, Macdonald, 2015).

Much research has been devoted to understanding the acceptance of vaccines. For instance, scholars have used quantitative and qualitative methods to explore vaccine decisions (Lama et al., 2020; Dubé et al., 2018; Marcu et al., 2015). Hesitancy to get vaccinated can be caused by a broad range of factors, including the compulsory nature of vaccines, unfamiliarity with vaccine-preventable diseases, and lack of trust in corporations and public health organizations (Salmon et al., 2015). Previous work has also indicated differences in vaccine acceptance across cultural and ethnic groups (Determann et al., 2016). Regarding the 2009 H1N1 pandemic, scholars have identified somefactors influencing vaccine decisions within the general population (Quinn et al., 2009) and healthcare workers (Wong et al., 2010).

Widespread vaccination can result in herd immunity, under which a specific population becomes resistant to the disease if a majority of its members develop immunity against it. Vaccine refusal could lead to a longer-standing disease, causing many more deaths than if vaccination was widespread. Therefore, a COVID-19 vaccine is urgently needed to overcome the crisis in the pandemic situation.

AI has entered the field of medicine, and much of the related research has been dedicated to the diagnosis and treatment of diseases (Oh et al., 2018; Li et al, 2014). Regarding the COVID-19 pandemic, AI is currently being used to diagnose cases from CT scans (Zheng et al., 2020), predict the spread of the virus (Yang et al., 2020), and in many other areas. For instance, an AI-based search system has identified an existing drug that could combat the disease (Richardson et al., 2020).

AI is also being employed in vaccine development. For instance, the Search Algorithm for Ligands (SAM), which is a program designed to identify compounds to improve the human immune system, has independently developed a highly effective flu vaccine (Masige, 2019). While AI could shorten the drug discovery process and improve the global health system by identifying new and more effective vaccines, it could also arguably contribute to the spread of anti-vaccine movements. The general public might develop trust in AI systems differently than they develop trust in their human counterparts. Therefore, employing AI in vaccine development might increase vaccine hesitancy and refusal, making it more difficult to fight against vaccine-preventable diseases.

## 2.3. Issues of Responsibility for Artificial Intelligence's Actions

The deployment of AI in various sectors of society has raised numerous ethical, legal, and moral questions. Researchers have proposed the concept of Responsible AI (Dignum, 2019), in which developers should take responsibility for all steps of developing and deploying their systems. However, scholars have also posited that doing so might not be viable given the unpredictability and complexity of self-learning and autonomous AI systems (Gunkel, 2017).

The concept of responsibility has various related meanings (Van de Poel, 2011), all of which have their attribution requirements (Van de Poel, Sand, 2018). Scholars have previously discussed whether AI and other stakeholders can satisfy these conditions, concluding that no entity fulfills all responsibility attribution requirements (Lima, Cha, 2020). Responsibility attribution for AI systems also suffers from the issue of 'many things' (Coeckelbergh, 2019): AI is a collection of various entities, technologies, and smaller interacting structures that makes the assignment of responsibility highly complex. In the context of medicine, scholars have shown that the public assigns responsibility and punishment for negative outcomes to both AI and human doctors, although to different extents (Lima et al., 2020).

Previous research has also proposed various moral and legal gaps arising from the deployment of AI systems. For instance, the responsibility gap is created by the lack of capacity to predict the behavior of self-learning machines by the manufacturer or operator (Matthias, 2004), which results in these entities not being able to be held morally and legally responsible for outcomes. Similarly, the accountability gap arises from a distance between the operator/manufacturer and the machine, making it difficult to attribute "causality to either the physical person or company that is behind the (electronic) agent" (Koops, Hildebrandt, Jaquet-Chiffelle, 2010). In medicine, an area where consequences are often a matter of life and death, successfully ascribing responsibility for mistakes and adverse outcomes is essential for developing safe and trustworthy systems.

# 3. Methods

## 3.1. Study Design

We conducted a cross-sectional study with a between-subjects design in which participants were presented two different news-like vignettes out of nine combinations (3X3).

## 3.2. Participants and Recruitment

We used quota sampling methods to recruit adults in the United States (US) and United Kingdom (UK) in mid-April 2020 after obtaining approval from the first author's Institutional Review Board. We used the Prolific crowdsourcing platform to recruit samples representative of current sex, age, and ethnicity demographic distributions in the US and UK. These countries were chosen because of the large number of cases and deaths in these countries at the time of the study.

We used partial eta2, which is a measure of effect size widely used with analyses of variance (ANOVAs) to calculate the sample size. To detect a small-to-moderate effect size (partial=0.03) with a power of .95 at the significance level of .05, our study design required sample size of 204 participants per country. We thus initially recruited 630 participants. Inclusion criteria were adults over 18 years. After excluding participants who failed the attention check question or did not spend enough time reading the given news-like vignettes during the survey, 572 survey answers were included in the data analysis.

## 3.3. Data Collection

The survey started by briefly introducing COVID-19 and the number of confirmed cases and deaths at the time of the study. The participants were then asked three questions addressing whether they had any close contact with suspected or confirmed patients and a set of concern-related questions to evaluate their level of concern regarding COVID-19.

We developed two sets of news-like vignettes. Two scholars that have expertise in fake news research reviewed the content and presentation of the news-like vignettes. Participants were randomly selected to view one pair of news-like vignettes. The first set of news-like vignettes introduced the launch of a vaccine trial against the novel coronavirus. We designed three versions of articles that differed solely in terms of who led to the development of the vaccine (see Appendix 1). One vignette ('human-developed') described a vaccine developed entirely by human researchers in collaboration with various institutes (n=202). Another vignette ('human-AI collaboration') introduced SAM, the AI program responsible for the first AI-developed vaccine (Masige, 2019) (n=180). This article described SAM as the entity that identified a set of compounds that seemed effective against COVID-19. Working alongside SAM, human researchers distinguished the best compound and synthesized the vaccine. The final vignette ('AI-developed') presented the AI program (i.e., SAM) as the entity that found the most effective compound and explicitly stated that the human researchers only synthesized the vaccine (n=190). The participants were then asked how willing they were to vaccinate 1) themselves, 2) their children, and 3) their elders. All questions were presented in random order, and participants answered them using a slider in the range of -1 and 1.

The survey subsequently presented another randomly selected news-like vignette that described a future scenario in mid-2021 in which a COVID-19 vaccine had finally been approved and made available to the public. Each news-like vignette framed the vaccine's effectiveness or side effects differently (see Appendix 2). The "positive" vignette introduced the vaccine as highly effective in inducing immunity against the novel coronavirus (n=206). This article explicitly compared the effectiveness of this vaccine to the average effectiveness of existing vaccines (WHO, 2020). On the other hand, the "negative" vignette discussed how there had been some reports of mild side effects, such as headaches and fever, in recipients of this new vaccine (n=168). This vignette stated that no complications had been reported and that the vaccine was the most effective way to combat the disease. The third was a "control" vignette, which excluded any mention of the vaccine's effectiveness or side effects (n=198).

After being shown one of the three vignettes in the second round, the participants were asked how they would assign responsibility and awareness to the various entities involved in developing and deploying the vaccine. Participants assigned to the human-developed vignette were presented with three entities to which they were asked to attribute these variables: the researchers who developed the vaccine, the government, and the health care worker administering the vaccine. In the human-AI collaboration or the AI-developed conditions, the respondents were additionally asked to assign responsibility and awareness to SAM, the company that developed it, and its main programmer.

We used the concept of credit (i.e., praise) and blame as measures of responsibility attribution. Participants who read the positive and control news-like vignettes were asked to attribute credit and awareness regarding the effectiveness and development of the COVID-19 vaccine to all entities above, respectively. Those who read a negative news-like vignette were asked about the levels of blame, awareness, and punishment they would assign to all entities for the vaccine's side effects. Entities were presented one at a time and in random order to all participants. The responses were recorded on a scale from 0 to 1 using a slider. Finally, participants were asked the same vaccine willingness-related questions from the previous news-like vignette.

## 3.4. Data Analysis

We analyzed our data by conducting ANOVAs to address statistically significant differences across groups using the lme4 library for the R programming language. We calculated least-square means and made pairwise comparisons between groups (with Tukey adjustments) using the lsmeans library for R. All models included a random-effects intercept to account for multiple measurements across participants (e.g., the participants indicated their level of vaccine acceptance for themselves, their children, and their elders). Interaction terms between the main variables were included in all tests. We employed a significance level threshold of .05. All plots present the group's least-square means and their respective 95% confidence intervals.

## 4. Results

### 4.1. Demographic Characteristics

The demographic characteristics of our samples are described in Table 1. They were divided by gender, age group, and education level. Female participants accounted for 49.8% and 48.4% of the samples in the US and UK, respectively. Most participants in both countries held either high school diplomas or bachelor's degrees.

Table 1. Demographics(N=572)

| Demographic Attributes | | US (n=287) | UK (n=285) |
|---|---|---|---|
| Gender | Male | 143 (48.5%) | 146 (51.2%) |
| | Female | 139 (49.8%) | 138 (48.4%) |
| | Other | 5 (1.7%) | 1 (0.4%) |
| Age | 18-24 yrs. old | 43 (15.0%) | 11 (3.9%) |
| | 25-34 yrs. old | 57 (19.9%) | 50 (17.5%) |
| | 35-49 yrs. old | 69 (24.0%) | 57 (20.0%) |
| | 49-64 yrs. old | 87 (30.3%) | 133 (46.7%) |
| | 65+ yrs. old | 31 (10.8%) | 34 (11.9%) |
| Education | High school | 113 (39.4%) | 117 (41.1%) |
| | Bachelor's degree | 133 (46.3%) | 114 (40.0%) |
| | Graduate school or higher | 41 (14.3%) | 54 (18.9%) |

## 4.2. Participants' Level of Concern about COVID-19

To evaluate the level of concern about COVID-19, we used four questions designed to quantify people's concerns regarding COVID-19 and their perceived likelihood of infection. The participants reported their level of concern in the range of -1 and 1 using a slider. The proposed scale had an acceptable level of internal consistency (Cronbach's=0.71). We divided participants into three quantiles of personal concern based on their responses to the concern-related questions. We categorized participants into three groups: those less concerned about the disease (little concern), those very concerned (much concern), and those in between (moderate concern). Figure 1 shows the cumulative distribution of responses and their categorization.

To evaluate the degree of medical use, we used a medical maximizer-minimizer scale (MMMS; Scherer et al., 2016) with a slider range of -1 and 1. Responses to all ten questions showed high internal consistency (Cronbach's α=0.86). Participants were divided into three categories: participants with a value of less than -0.333 were categorized as medical minimizers, those with a value greater than –0.333 and less than 0.333 were in between, and participants with a value equal to or greater than 0.333 were considered medical maximizers.

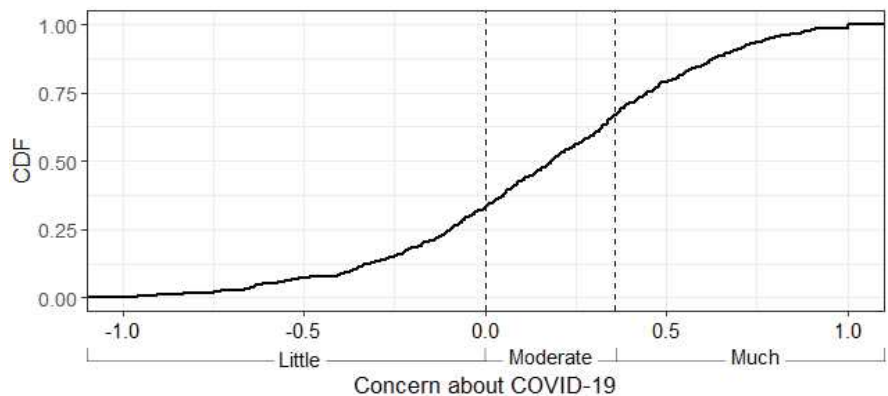

Figure 1. Cumulative distribution of average personal concern regarding COVID-19(N=572)

### 4.3. Willingness to Get Vaccinated against COVID-19

Vaccine acceptance based on the participants' concerns and healthcare usage is described in Figure 2. The participants in this study were marginally predisposed to accept a vaccine against the novel coronavirus, with a reported median and mean willingness of 0.170 and 0.143, respectively. We observed a lower overall willingness to vaccinate children (F-test, p<.001) regardless of the participants' level of concern regarding COVID-19 or their predisposition to seek health care. Participants revealed a lower acceptance of the vaccine for children (95% CI [-0.083, 0.036]) compared to their willingness to vaccinate themselves ([0.148, 0.268]) and their elders ([0.150, 0.270]). However, no difference was found between their inclination to get personally vaccinated against COVID-19 and the predisposition to take their elders to get the vaccine. We did not find any significant differences between the US and UK participants (F-test, p=.67).

The level of concern regarding the novel coronavirus correlated positively with the overall willingness to accept a newly developed COVID-19 vaccine for all recipients (see Figure 2a, F-test, p<.001). Those more concerned about the disease were more predisposed to get vaccinated ([0.171, 0.359]) than those less concerned about it ([-0.090, 0.102]). Those with moderate concerns did not indicate significantly different vaccine acceptance levels from the other two groups ([0.032, 0.215]). Figure 2b shows that medical maximizers also reported a higher willingness to get vaccinated (F-test, p<.05, [0.125, 0.331]) than medical minimizers ([-0.078, 0.150]). Those in the center of the MMMS scale ([0.078, 0.183]) did not differ from the other two groups.

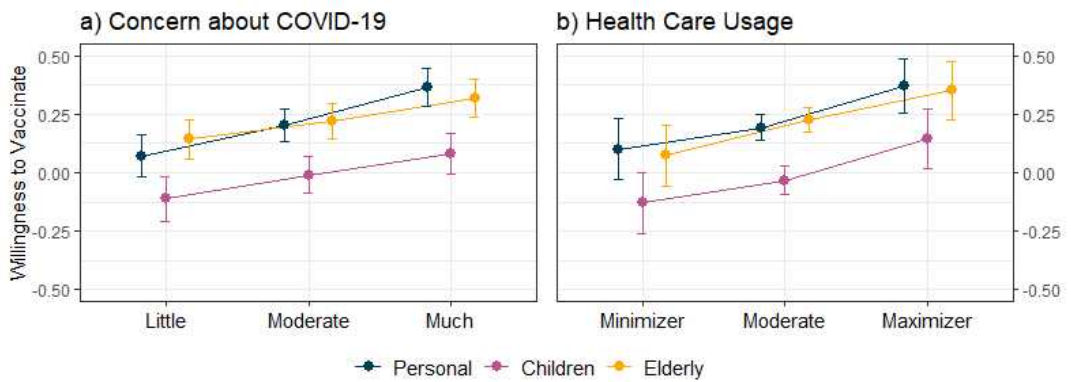

Figure 2. Willingness to accept a COVID-19 vaccine according to the recipient (N=572)

The differences in vaccination acceptance depending on the participants' demographics, previous experiences with vaccines, and contact with COVID-19 were also examined. Figure 3 shows the willingness to accept a COVID-19 vaccine by sex (see Figure 3a), previous experiences with vaccines (see Figure 3b), and whether there had been COVID-19 cases in the respondent's city (see Figure 3c). Female participants were on average more hesitant (F-test, $p<.05$, [-0.086, 0.160]) about the COVID-19 vaccine than male participants ([0.029, 0.278]). The participants who had received all their countries' required vaccines reported higher acceptance (F-test, $p<.001$, [0.095, 0.450]) than those who had received only some ([-0.086, 0.293]) or none ([-0.098, 0.318]). Last, those residing in cities where COVID-19 had infected at least one resident at the time of the study also reported less hesitancy towards the vaccine (F-test, $p<.05$).

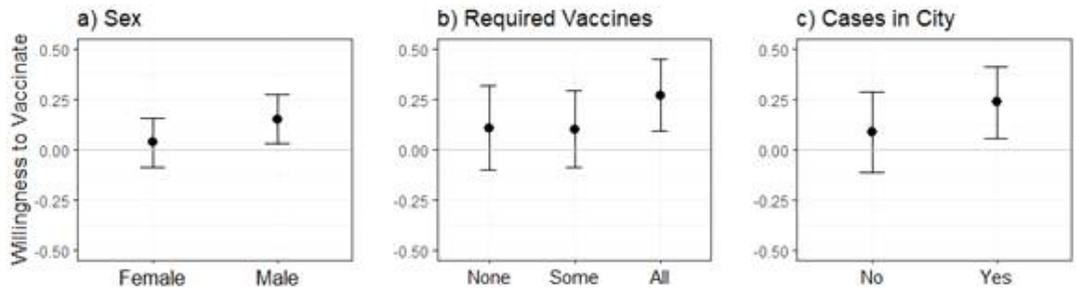

Figure 3. Acceptance of the COVID-19 vaccine according to sex, vaccination history, and cases of COVID-19 in the city of residence(N=572)

Figure 4 shows the mean vaccine acceptance scores regardless of the recipient and depending on the entity that led its development. The participants' willingness to get vaccinated was not significantly altered by the involvement of AI in the development of vaccines (F-test, $p=.096$). We also did not observe any significant interaction of this variable with the participants' level of concern about COVID-19 ($p=.894$) or their health care usage ($p=.148$).

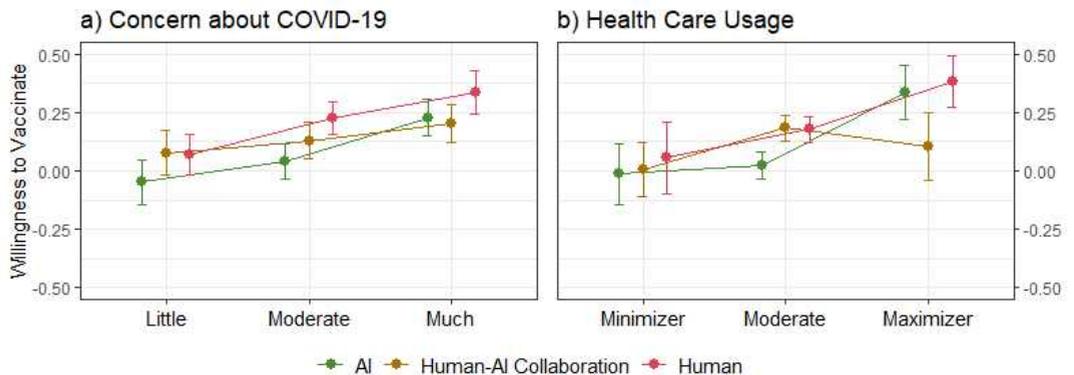

Figure 4. Willingness to accept a COVID-19 vaccine depending on the recipient (N=572)

## 4.4. Factors that Influenced the Participants' Willingness to Receive a COVID-19 Vaccination

Table 2 describes the influence of stimuli introducing the vaccine's effectiveness and side effects on participants' acceptance of a COVID-19 vaccine. The approach used to introduce the consequences of the COVID-19 vaccine affected how much people modified their initial acceptance of vaccination (F-test, p<.001). More positive framing of the consequences of the COVID-19 vaccine increased the overall willingness to get vaccinated (95% CI of willingness change [0.204, 0.311]) compared to the control treatment ([0.042, 0.149]). Interestingly, a more negative framing also resulted in a more positive willingness change than the control stimulus ([0.138, 0.253]).

The news stimuli increased participants' willingness to vaccinate children (F-test, p<.001, [0.175,0.253]) to a greater degree than in the case of personal vaccination ([0.133,0.210]) or vaccination of elderly persons ([0.124,0.202]). Finally, the participants' initial willingness was strongly correlated with how much they modified their inclination to get vaccinated upon receiving the intervention (F-test, p<.001). Participants in the lower tercile of vaccine acceptance reported a higher willingness change ([0.324, 0.411]) than those in both the center ([0.144, 0.229]) and upper ([-0.048, 0.038]) terciles of willingness to get vaccinated against COVID-19. Those who reported an average willingness were also more affected by the vignette than respondents in the upper tercile.

Table 2. Participants' willingness change upon presentation of news stimuli highlighting the vaccine's effectiveness, side effects, or neither(i.e., control condition).

| | | Contrast | p-value |
|---|---|---|---|
| Consequences | Effectiveness-Control | 0.162 | <.001 |
| | Side Effects-Control | 0.100 | <.001 |
| | Side Effects-Effectiveness | -0.062 | .229 |
| Recipient | Child-Personal | 0.042 | .006 |
| | Child-Elderly | 0.051 | <.001 |
| | Personal-Elderly | 0.009 | .795 |
| Quantile of willingness | Lower Tercile-Higher Tercile | 0.373 | <.001 |
| | Middle Tercile-Higher Tercile | 0.192 | <.001 |
| | Middle Tercile-Lower Tercile | -0.181 | <.001 |

## 4.5. Responsibility for the COVID-19 Vaccination

Figure 5 shows the level of blame, credit, and awareness participants ascribed to all entities addressed in our study based on the positive, negative, and control news stimuli they were shown. Even though liability could be considered one of the meanings of responsibility (Van de Poel, Sand, 2018), we treat blame and praise (i.e., credit) as responsibility and punishment as liability. The study participants assigned substantially less blame for the vaccine's side effects than credit for its development and effectiveness across all entities (F-test, p<.001). This decrease in responsibility appeared to a lesser extent in the case of the government. The overall responsibility ascribed to AI was similar to that attributed to its programmer. The AI program was assigned marginally less blame and credit than its human counterpart.

The level of awareness ascribed to all entities was overall lower in the case of negative framing of the consequences of the vaccine (F-test, p<.001). Therefore, people believed entities were less aware of the possible side effects of a vaccine than of its effectiveness (i.e., positive news stimuli) and its development in general (i.e., control news stimuli). The level of awareness assigned to AI was lower than that assigned to most entities.

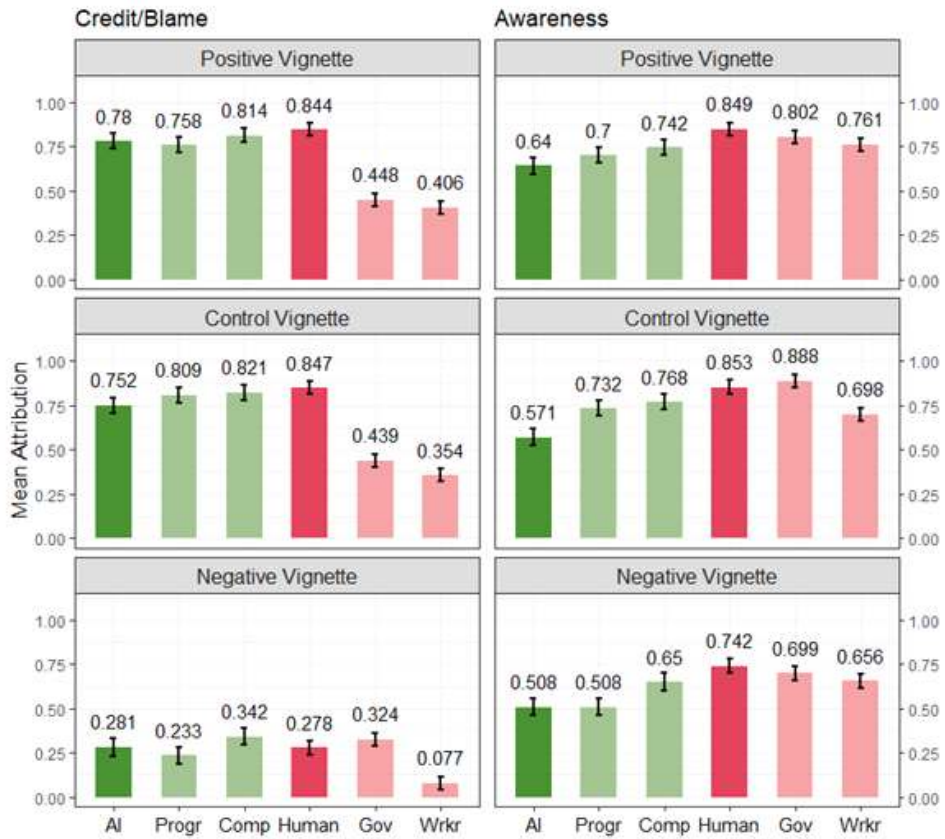

Figure 5. Assignment of blame, credit, and awareness to all entities involved in the development and deployment of the COVID-19 vaccine depending on how the consequences of the vaccine were framed(N=572).

* AI=AI system, Progr=Programmer, Comp=Company, Human=Human researchers, Gov=Government, Wrkr=Health care worker

Figure 6 shows to what extent participants assigned punishment for the side effects of the vaccine. Participants were asked these questions only if they had been assigned a news-like vignette that excluded any mentions of effectiveness (i.e., the negative stimuli). The overall punishment assigned to all entities was substantially lower than the responsibility assigned. The government was ascribed higher liability than all entities, except for the AI-developing company (all p<.05).

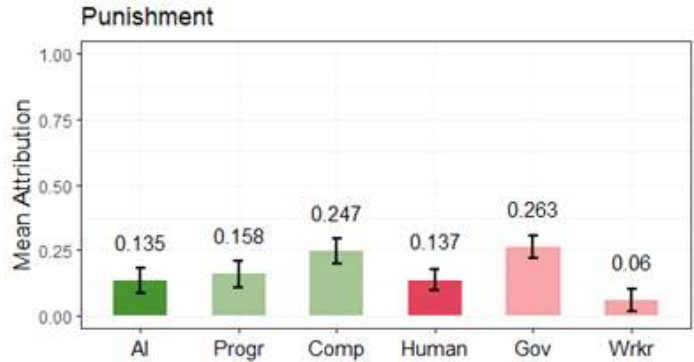

Figure 6. Assignment of punishment to all entities involved in developing the COVID-19 vaccine(N=572).

## 5. Discussion

Our results indicate that people who are less concerned about the disease, who have a history of vaccination hesitancy or refusal, and who are less affected by the virus are less likely to get vaccinated against COVID-19. Governments could develop health policies to promote vaccination against the novel coronavirus based on this finding. For example, national campaigns that increase awareness of the risk and side effects of COVID-19 might increase perceived susceptibility and severity. These efforts could also increase the acceptance of the vaccine and compliance with preventive measures, such as social distancing if targeted at those less concerned.

People were less willing to vaccinate their children than themselves or their elders. While our results indicate that people are aware that the oldest in society must also be vaccinated against the novel coronavirus, people hesitate to vaccinate the youngest. This tendency may be attributed to misinformation that children are resistant or even immune to COVID-19 (Powder, 2020). These rumors have been disseminated by the general population, world leaders, and public figures in various countries (Bello, 2020; O'Kane, 2020). Those who defend more lenient quarantine policies often argue that society's youngest stratum should not be isolated due to the lower risk of complications. However, defenders of these proposals neglect the possibility of spread from the youngest to the elderly, i.e., those more susceptible to complications due to COVID-19. Moreover, the burden of disease caused by the long-term impact of the coronavirus (Al-Aly et al., 2021), such as cardiovascular complications (Xiong et al., 2020) and mental health issues (Kumar, Nayar, 2020), could decrease the quality of life and increase future healthcare costs of children.

Our findings suggest that government and health organizations, such as the national Centers for Disease Control and Prevention (CDC) and the WHO, should highlight the importance of vaccinating the youngest groups in society against COVID-19 as a form of control of the disease once a vaccine is available to the population. In addition, we propose investigating in future studies why the public, especially parents, are reluctant to vaccinate their children. Studies in developed countries examining the factors that lead parents to accept or reject different vaccines for their children have revealed that various factors, such as knowledge/information about vaccination, past experiences with vaccination services, health professionals' recommendations, use of complementary and alternative medicine, risk perceptions, and trust influence parents' decisions (Dubé et al., 2013; Benin et al., 2006). Therefore, it is necessary to investigate why parents are reluctant to receive this specific vaccine for their children.

Including AI systems in the development of vaccines does not seem to affect vaccine acceptance. Our results indicate that an AI-developed vaccine has a similar acceptance rate as those developed by human researchers. AI might shorten drug discovery time and improve the global health system (Masige, 2019). Therefore, the inclusion of AI in medicine should not contribute to vaccine hesitancy and refusal while advancing vaccine development without much public backlash. A crucial future line of work is understanding how incorporating AI in medicine could lead to the creation of misinformation, thereby hindering vaccine acceptance. For instance, the COVID-19 has led to the spread of false technology-related claims, e.g., that the vaccine was being developed for implanting tracking microchips, and similar rumors might emerge about AI (e.g., "this vaccine will turn you into AI").

Our results suggest that even if side effects are communicated to the general public, vaccine acceptance might increase depending on how effective the vaccine is reported to be. Therefore, promotion campaigns must be drafted carefully to address public concerns while promoting the vaccine's effectiveness against the disease. It is important to highlight the ethical and moral obligation to communicate the possible side effects and risks of a vaccine to the general public. Governments should not conceal any information regarding the potential risks of getting vaccinated.

Participants in this study assigned a high level of credit to the AI system for the vaccine's development and effectiveness. Moreover, in terms of blame, AI was attributed similar levels to those attributed to human researchers for the vaccine's side effects. In agreement with previous work (Lima et al., 2020), the participants attributed responsibility to AI systems for their actions, although they were not considered aware of their actions. Even though a marginally smaller degree of responsibility was attributed to AI than to their human counterparts, their blame or credit for the vaccine outcomes was similar to those of their main programmers.

The government and the human researchers involved in developing the vaccine were ascribed the highest degree of awareness by the participants. Even though the government was assigned less credit for the vaccine's development and effectiveness, it was attributed one of the highest levels of blame. Alongside the AI-developing company, the government was assigned the most punishment for the vaccine's side effects. Therefore, our results indicate a public attribution of responsibility for vaccination to the government, especially if the vaccine has any side effects. This emphasizes the role of the government in testing and approving safe and effective treatments for the current pandemic.

Our results suggest that people attribute responsibility for an AI system's actions to the system and other entities involved in its development and deployment. This supports a "extended agency theory," where responsibility and moral agency are jointly distributed across both human and AI systems, a previously proposed response to the various responsibility gaps raised by AI deployment (Gunkel, 2017).

## 6. Conclusion

The current study indicates four main findings that suggest how governments might want to focus their COVID-19 vaccination policies in the future. First, the participants exhibited a lower level of willingness to vaccinate children than themselves or their elders. Second, those less concerned about the disease or more doubtful about seeking healthcare were significantly more hesitant about the vaccine. Third, people with a previous history of vaccination hesitancy and communities less affected by the virus also reported a lower willingness to accept the COVID-19 vaccine. Finally, our results suggest that promotion campaigns highlighting the vaccine's effectiveness against the pandemic can increase willingness to get vaccinated, especially among those initially more hesitant about vaccinations.

This study highlights the government's crucial role in ensuring the safety of the COVID-19 vaccines even during the early stage of the virus. Our results suggest that tailored health policies to increase perceived susceptibility and perceived severity could increase the number of people willing to accept a COVID-19 vaccine. These campaigns could promote vaccination among children as people are less willing to vaccine the youngest in society against the novel coronavirus. A possibility could be to focus these promotion efforts on highlighting the vaccine's effectiveness against the pandemic in a consistent voice. We underscore that any campaigns adopted by governments should be transparent regarding vaccines' effectiveness and side effects by disseminating accurate scientific knowledge in a comprehensible manner. Scientific information might not be approachable, and authorities must play a role in correctly informing the public. The findings presented in this study capturing the public's perceptions during the early stage of the pandemic could be valuable for better understanding public perceptions and concerns about a novel disease.

Appendix 1.

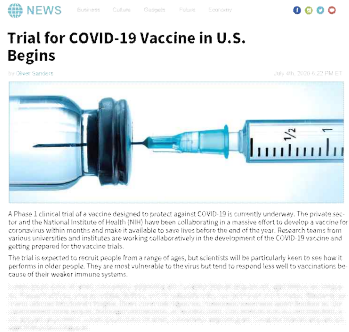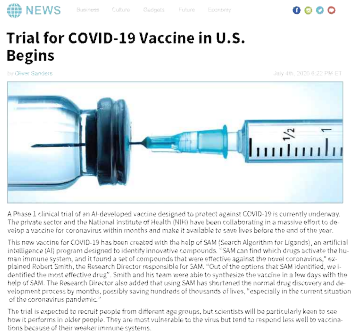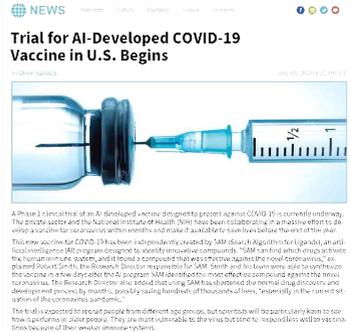

Human-developed vaccine        Human-AI collaboration        AI-developed vaccine

Appendix 2.

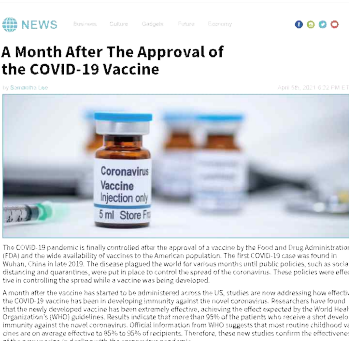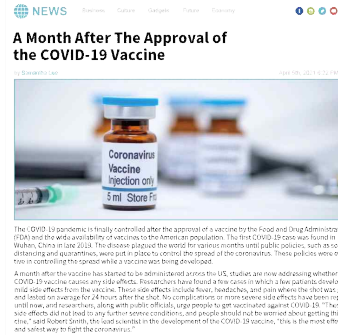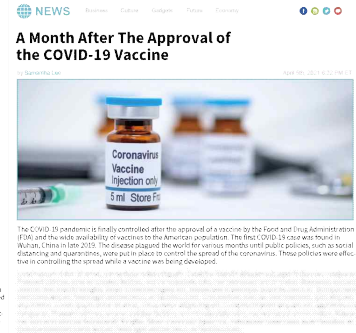

Positive framing introducing the high effectiveness of the vaccine        Negative framing describing reports of side effects due to the vaccine        Control framing which does not explicitly discuss the vaccine's effectiveness or side effects